\documentclass[aps,prl,twocolumn,showpacs,superscriptaddress]{revtex4}  	
\usepackage{graphicx}  	
\usepackage{dcolumn}   	
\usepackage{bm}        	
\usepackage{amssymb}   
\usepackage{multirow}
\usepackage{color}
\usepackage{units}

\newcommand{\etal}{{\it et al.}}

\newcommand{\delmsq}[1]{\ensuremath{\Delta m^2_{ #1 }}}

\begin{document}


\leftline{FERMILAB-PUB-11-040-PPD  \& BNL-94498-2010-JA; \hskip1.5in submitted to PRL; \today} 

\title
{
Measurement of the neutrino mass splitting and flavor mixing by MINOS
}





\newcommand{\Berkeley}{Lawrence Berkeley National Laboratory, Berkeley, California, 94720 USA}
\newcommand{\Cambridge}{Cavendish Laboratory, University of Cambridge, Madingley Road, Cambridge CB3 0HE, United Kingdom}
\newcommand{\FNAL}{Fermi National Accelerator Laboratory, Batavia, Illinois 60510, USA}
\newcommand{\RAL}{Rutherford Appleton Laboratory, Science and Technologies Facilities Council, OX11 0QX, United Kingdom}
\newcommand{\UCL}{Department of Physics and Astronomy, University College London, Gower Street, London WC1E 6BT, United Kingdom}
\newcommand{\Caltech}{Lauritsen Laboratory, California Institute of Technology, Pasadena, California 91125, USA}
\newcommand{\Alabama}{Department of Physics and Astronomy, University of Alabama, Tuscaloosa, Alabama 35487, USA}
\newcommand{\ANL}{Argonne National Laboratory, Argonne, Illinois 60439, USA}
\newcommand{\Athens}{Department of Physics, University of Athens, GR-15771 Athens, Greece}
\newcommand{\NTUAthens}{Department of Physics, National Tech. University of Athens, GR-15780 Athens, Greece}
\newcommand{\Benedictine}{Physics Department, Benedictine University, Lisle, Illinois 60532, USA}
\newcommand{\BNL}{Brookhaven National Laboratory, Upton, New York 11973, USA}
\newcommand{\Cleveland}{Cleveland Clinic, Cleveland, Ohio 44195, USA}
\newcommand{\Delhi}{Department of Physics \& Astrophysics, University of Delhi, Delhi 110007, India}
\newcommand{\GEHealth}{GE Healthcare, Florence South Carolina 29501, USA}
\newcommand{\Harvard}{Department of Physics, Harvard University, Cambridge, Massachusetts 02138, USA}
\newcommand{\HolyCross}{Holy Cross College, Notre Dame, Indiana 46556, USA}
\newcommand{\IIT}{Physics Division, Illinois Institute of Technology, Chicago, Illinois 60616, USA}
\newcommand{\Iowa}{Department of Physics and Astronomy, Iowa State University, Ames, Iowa 50011 USA}
\newcommand{\Indiana}{Indiana University, Bloomington, Indiana 47405, USA}
\newcommand{\ITEP}{High Energy Experimental Physics Department, ITEP, B. Cheremushkinskaya, 25, 117218 Moscow, Russia}
\newcommand{\JMU}{Physics Department, James Madison University, Harrisonburg, Virginia 22807, USA}
\newcommand{\LASL}{Nuclear Nonproliferation Division, Threat Reduction Directorate, Los Alamos National Laboratory, Los Alamos, New Mexico 87545, USA}
\newcommand{\Lebedev}{Nuclear Physics Department, Lebedev Physical Institute, Leninsky Prospect 53, 119991 Moscow, Russia}
\newcommand{\LLL}{Lawrence Livermore National Laboratory, Livermore, California 94550, USA}
\newcommand{\LosAlamos}{Los Alamos National Laboratory, Los Alamos, New Mexico 87545, USA}
\newcommand{\MIT}{Lincoln Laboratory, Massachusetts Institute of Technology, Lexington, Massachusetts 02420, USA}
\newcommand{\Minnesota}{University of Minnesota, Minneapolis, Minnesota 55455, USA}
\newcommand{\Crookston}{Math, Science and Technology Department, University of Minnesota -- Crookston, Crookston, Minnesota 56716, USA}
\newcommand{\Duluth}{Department of Physics, University of Minnesota -- Duluth, Duluth, Minnesota 55812, USA}
\newcommand{\Ohio}{Center for Cosmology and Astro Particle Physics, Ohio State University, Columbus, Ohio 43210 USA}
\newcommand{\Otterbein}{Otterbein College, Westerville, Ohio 43081, USA}
\newcommand{\Oxford}{Subdepartment of Particle Physics, University of Oxford, Oxford OX1 3RH, United Kingdom}
\newcommand{\PennState}{Department of Physics, Pennsylvania State University, State College, Pennsylvania 16802, USA}
\newcommand{\PennU}{Department of Physics and Astronomy, University of Pennsylvania, Philadelphia, Pennsylvania 19104, USA}
\newcommand{\Pittsburgh}{Department of Physics and Astronomy, University of Pittsburgh, Pittsburgh, Pennsylvania 15260, USA}
\newcommand{\IHEP}{Institute for High Energy Physics, Protvino, Moscow Region RU-140284, Russia}
\newcommand{\Rochester}{Department of Physics and Astronomy, University of Rochester, New York 14627 USA}
\newcommand{\RoyalH}{Physics Department, Royal Holloway, University of London, Egham, Surrey, TW20 0EX, United Kingdom}
\newcommand{\Carolina}{Department of Physics and Astronomy, University of South Carolina, Columbia, South Carolina 29208, USA}
\newcommand{\SLAC}{Stanford Linear Accelerator Center, Stanford, California 94309, USA}
\newcommand{\Stanford}{Department of Physics, Stanford University, Stanford, California 94305, USA}
\newcommand{\StJohnFisher}{Physics Department, St. John Fisher College, Rochester, New York 14618 USA}
\newcommand{\Sussex}{Department of Physics and Astronomy, University of Sussex, Falmer, Brighton BN1 9QH, United Kingdom}
\newcommand{\TexasAM}{Physics Department, Texas A\&M University, College Station, Texas 77843, USA}
\newcommand{\Texas}{Department of Physics, University of Texas at Austin, 1 University Station C1600, Austin, Texas 78712, USA}
\newcommand{\TechX}{Tech-X Corporation, Boulder, Colorado 80303, USA}
\newcommand{\Tufts}{Physics Department, Tufts University, Medford, Massachusetts 02155, USA}
\newcommand{\UNICAMP}{Universidade Estadual de Campinas, IFGW-UNICAMP, CP 6165, 13083-970, Campinas, SP, Brazil}
\newcommand{\UFG}{Instituto de F\'{i}sica, Universidade Federal de Goi\'{a}s, CP 131, 74001-970, Goi\^{a}nia, GO, Brazil}
\newcommand{\USP}{Instituto de F\'{i}sica, Universidade de S\~{a}o Paulo,  CP 66318, 05315-970, S\~{a}o Paulo, SP, Brazil}
\newcommand{\Warsaw}{Department of Physics, Warsaw University, Ho\.{z}a 69, PL-00-681 Warsaw, Poland}
\newcommand{\Washington}{Physics Department, Western Washington University, Bellingham, Washington 98225, USA}
\newcommand{\WandM}{Department of Physics, College of William \& Mary, Williamsburg, Virginia 23187, USA}
\newcommand{\Wisconsin}{Physics Department, University of Wisconsin, Madison, Wisconsin 53706, USA}
\newcommand{\deceased}{Deceased.}

\affiliation{\ANL}
\affiliation{\Athens}
\affiliation{\BNL}
\affiliation{\Caltech}
\affiliation{\Cambridge}
\affiliation{\UNICAMP}
\affiliation{\FNAL}
\affiliation{\UFG}
\affiliation{\Harvard}
\affiliation{\HolyCross}
\affiliation{\IIT}
\affiliation{\Indiana}
\affiliation{\Iowa}
\affiliation{\UCL}
\affiliation{\Minnesota}
\affiliation{\Duluth}
\affiliation{\Otterbein}
\affiliation{\Oxford}
\affiliation{\Pittsburgh}
\affiliation{\RAL}
\affiliation{\USP}
\affiliation{\Carolina}
\affiliation{\Stanford}
\affiliation{\Sussex}
\affiliation{\TexasAM}
\affiliation{\Texas}
\affiliation{\Tufts}
\affiliation{\Warsaw}
\affiliation{\WandM}

\author{P.~Adamson}
\affiliation{\FNAL}

\author{C.~Andreopoulos}
\affiliation{\RAL}


\author{R.~Armstrong}
\affiliation{\Indiana}

\author{D.~J.~Auty}
\affiliation{\Sussex}


\author{D.~S.~Ayres}
\affiliation{\ANL}

\author{C.~Backhouse}
\affiliation{\Oxford}




\author{G.~Barr}
\affiliation{\Oxford}









\author{M.~Bishai}
\affiliation{\BNL}

\author{A.~Blake}
\affiliation{\Cambridge}


\author{G.~J.~Bock}
\affiliation{\FNAL}

% ...

\collaboration{The MINOS Collaboration}
\noaffiliation




\date{\today}


\begin{abstract}

Measurements of neutrino oscillations  using the disappearance of muon neutrinos from the Fermilab NuMI neutrino beam as observed by the two MINOS detectors are reported.
New analysis methods have been applied to an enlarged data sample from an exposure of  $7.25 \times 10^{20}$ protons on target.
A fit to neutrino oscillations yields values of $|\Delta m^2| = (2.32^{+0.12}_{-0.08})\times10^{-3}$\,eV$^2$ for the atmospheric mass splitting and $\rm \sin^2\!(2\theta)  > 0.90$ (90\%\,C.L.) for the mixing angle. 
Pure neutrino decay and quantum decoherence hypotheses are excluded at $7$ and $9$ standard deviations, respectively.
  
\end{abstract}


\maketitle



Neutrino masses and flavor mixing influence the role of neutrinos in fundamental physics processes~\cite{ref:Astrophysics}  and may point to the mechanism that gives rise to the matter-antimatter asymmetry in the Universe~\cite{ref:Leptogenesis}.
A variety of phenomena observed with neutrinos originating in the Earth's atmosphere or the Sun and those produced by nuclear reactors or accelerators can be described consistently  by  quantum-mechanical mixing of the weak flavor states of neutrinos.
The underlying mechanism of neutrino mixing resulting in neutrino oscillations, well established by several experiments over the last decade~\cite{ref:SuperK,ref:SNO,ref:K2K,ref:KamLAND,ref:MINOS}, is governed by the $3\times3$ unitary PMNS matrix~\cite{ref:PMNS}, which can be parametrized using three mixing angles and a CP-violating phase.
Evolution of neutrino flavor eigenstates in vacuum depends additionally on the ratio of the distance traveled to the neutrino energy ($L/E$) and the splitting  between the squared masses of neutrino mass eigenstates $i$ and $j$,  $\Delta m^2_{ji}=m^2_j-m^2_i$.
For three neutrino mass eigenstates there are two independent mass splittings.
MINOS, a long-baseline experiment with L/E =  $\mathcal{O}(500$\,km/GeV),  is sensitive to the larger (atmospheric)
mass splitting through the disappearance of muon neutrinos~\cite{ref:dm2note}.


The MINOS experiment uses two detectors separated by a distance  of \unit{734}\,km, both  placed in the intense NuMI neutrino beam from Fermilab~\cite{ref:NuMI-beam}.
The Near Detector is used primarily to characterize the NuMI beam near its production.
The Far Detector measures the event rate and energy spectra after the neutrinos have traveled  through the Earth's crust. 
In an earlier publication~\cite{ref:MINOS},  MINOS  presented the most precise measurement to date of the atmospheric mass splitting using data from a beam exposure of $3.36 \times 10^{20}$ protons on target (POT).
The results in this letter are based on an exposure of \unit{$7.25\times10^{20}$}\,POT,  involve additional event categories, and employ an improved analysis methodology.
%


The NuMI beam~\cite{ref:NuMI-beam} operates with \unit{120}\,GeV/c protons directed onto a graphite target of two interaction lengths.
Positively charged hadrons produced in the target are focused towards the beam axis by two magnetic horns. 
The neutrino beam is  the product of pion, kaon, and muon decays occurring downstream of the target, primarily along a \unit{675}\,m long decay pipe,  evacuated for the first half of the data set~\cite{ref:MINOS}, but later filled with \unit{0.9}\,atm helium for structural reasons.
Data taken at different  relative horn-target longitudinal positions  and horn currents were used to tune the neutrino beam simulation~\cite{ref:MINOS-PRD-2008}. 
The effect of the helium in the decay pipe and an observed decrease in neutrino flux per POT, attributed to target degradation, are incorporated into the simulations.
Most of the data were taken in three run periods (Runs~I$-$III), with the target placed in the most downstream position, yielding an energy spectrum of neutrino interactions peaking at \unit{3}\,GeV.
A small amount of the data was taken with the target placed upstream in the high energy (HE) configuration which results in an energy spectrum that peaks at \unit{9}\,GeV.


Both MINOS  detectors~\cite{ref:MINOS-NIM-detectors,ref:Caldet} are  placed on the NuMI beam line axis.
Each is a tracking, sampling calorimeter, built of  \unit{2.54}\,cm thick iron plates interleaved with scintillator planes composed of  \unit{1}\,cm thick,  \unit{4.1}\,cm wide scintillator strips, arranged in two alternating orthogonal views and read out using multi-anode photomultipliers.
The Near Detector is located \unit{1.04}\,km downstream from the target and has a \unit{23.7}\,t fiducial mass. 
The Far Detector has a \unit{4.2}\,kt fiducial (\unit{5.4}\,kt total) mass.  
Both detectors are magnetized, with a toroidal magnetic field oriented to focus negatively charged particles.

 
In comparison to the previous analysis~\cite{ref:MINOS}, the data set has been substantially increased
and several changes  to the simulation, reconstruction, and analysis methodology have been introduced.
This analysis also benefits from the inclusion of new event samples: events with a reconstructed track of positive charge and events originating outside of the fiducial volume, including the surrounding rock.
The geometrical modeling of the target and focusing system~\cite{ref:Pavlovic} of the NuMI beam line, using  the FLUGG software package~\cite{ref:FLUGG}, combining  GEANT4~\cite{ref:GEANT} geometry with the FLUKA~\cite{ref:Fluka} hadron production, has improved the beam simulation.
As previously, the detector simulation uses GEANT3 with NEUGEN3~\cite{ref:NEUGEN} as the neutrino interaction generator.


The most significant analysis improvement is achieved by employing a k-nearest-neighbor (kNN) algorithm~\cite{ref:kNN-reference} for estimation of the energy of showers produced by low energy hadronic cascades. 
In MINOS, a charged current interaction of a muon neutrino is typically characterized by a muon track and a hadronic cascade, reconstructed as one or more showers.
The energy resolution of events with a hadronic cascade is usually limited by the calorimetric measurement of the showers, which has a resolution of $56\% /\sqrt{\rm E(GeV)} \oplus 2\%$~\cite{ref:MINOS-NIM-detectors}. %
This can be compared to a resolution of 4.6\% or 11\% for \unit{3}\,GeV muon tracks measured by range or curvature, respectively.
The new hadronic energy estimator uses three event characteristics: the summed reconstructed energy deposited by showers within {1}\,m of the track vertex; the sum of the energy in the largest two reconstructed showers associated with the event; and the length of the longest shower~\cite{ref:ChrisBPhD}.
The hadronic energy is taken as the mean true hadronic energy of the closest Monte Carlo events in this three dimensional feature space. 
Monte Carlo studies show that the new algorithm improves shower energy resolution from  55\%  to 43\% for showers between \unit{$1.0-1.5$}\,GeV.


The identification of charged current neutrino interactions uses the energy deposition and its fluctuation along a track to discriminate muon tracks from spurious tracks reconstructed from hadronic activity in neutral current interactions.
However, this identification method, used in the prior publication~\cite{ref:MINOS,ref:Rustem}, does not resolve events with muons crossing 10 or fewer detector planes (corresponding to about \unit{500}\,MeV of muon energy).
An additional procedure is applied to reclaim some of these events by constructing a new kNN classifier from the total pulse height of the last 5 scintillator strips associated with a track, and from two quantities that are correlated with track scattering~\cite{ref:JasminePhD}.  
The new selection achieves  a  90\%  charged current efficiency.
Below \unit{2}\,GeV, the efficiency is 77\% with 6.5\% neutral current contamination.

\begin{figure}
\begin{center}
\includegraphics[viewport= 0 30 567 384, keepaspectratio,width= 0.49 \textwidth, clip=true]{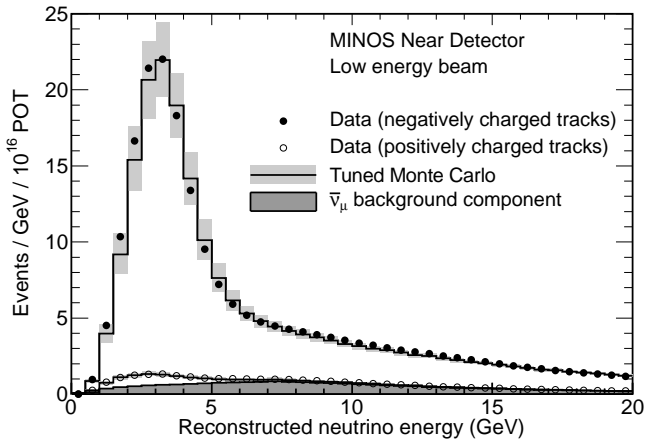}
\end{center}
\caption {
The energy spectra of fully reconstructed events in the Near Detector classified as
charged current interactions. The solid and open circles show the data
reconstructed with negative or positive track charge, respectively. The solid lines show the
tuned Monte Carlo with a shaded error band due to systematic uncertainties. The
shaded area at the bottom represents the simulated antineutrino component.
}
\label{fig:NDdata}
\end{figure}

Events classified as charged current interactions are used irrespective of the reconstructed charge-sign of the track. 
Compared to the previous analysis~\cite{ref:MINOS}, which used only events with a well identified negative track charge,  events at low energy, where track charge-sign reconstruction is less reliable, have been now recovered. 
Below \unit{6}\,GeV, the main oscillation energy range, 98.0\% of all selected events arise from neutrino interactions; the antineutrino component, shown in Fig.~\ref{fig:NDdata}, is small and contributes primarily at higher energies.
This antineutrino component is assumed to oscillate with the same parameters as the neutrinos.

The predicted energy spectrum in the Far Detector is calculated from the spectrum measured in the Near Detector, using a technique that takes account of the kinematics of neutrino production in the beam and of the geometry of the NuMI beam line~\cite{ref:MINOS-PRD-2008}. 
The Near Detector events with tracks of positive and negative reconstructed charges are used separately to provide energy spectra predictions at the Far Detector~\cite{ref:JustinThesis,ref:StephenPhD}. 
The Far Detector events with a reconstructed negative track charge are further divided into five quantiles based on energy resolution~\cite{ref:StephenPhD} determined by simulations and test beam measurements~\cite{ref:Caldet}. 
This division increases the sensitivity with which MINOS can measure the neutrino mass splitting and mixing~\cite{ref:JohnMarshall}, since events with the most precisely reconstructed energy carry the most precise information about the energy dependence of charged current event disappearance.


This analysis includes interactions originating in the rock and outside of the Far Detector fiducial volume.  
Such interactions are only partially reconstructed, and are characterized by the measured muon and its detector entry position~\cite{ref:Aaron,ref:MattPhD}.
The predicted energy spectrum for these events is derived using the same method as for the fully reconstructed events.
The partially and fully reconstructed samples have comparable statistics, but the partially reconstructed events contribute primarily to establishing the overall event rate since they are due to neutrinos that are not well measured and are predominantly at higher energies.


\begin {table}[b]
{
{
\begin {center}
\begin {tabular} {l c r }
\hline
Source   of 						& $\rm \delta (\delmsq{})$  	&  $\rm \delta (sin^2\!(2\theta))$	\\
systematic uncertainty	                 		& $\rm (10^{-3}\,eV^2)$       	&                                        			\\
\hline
(a) Hadronic energy 							&     0.051		&  $<$ 0.001	\\
(b) $\mu$ energy (range 2\%, curv. 3\%)          		&     0.047 	&         0.001	\\                       
(c) Relative normalization   $(1.6\%)$   			&     0.042		&  $<$ 0.001	\\
(d) NC contamination $(20\%)$             			&     0.005 	&         0.009	\\
(e) Relative hadronic energy  $(2.2\%)$	     		&     0.006  	&         0.004 	\\            
(f) $\rm \sigma_\nu(E_\nu < 10~GeV)$			&     0.020 	&         0.007 	\\
(g) Beam flux           							&     0.011 	&  	   0.001   	\\
(h) Neutrino-antineutrino separation                      	&     0.002  	&         0.002    	\\
(i) Partially reconstructed events				&     0.004		&	   0.003	\\
\hline
Total systematic uncertainty                               		&     0.085   	&         0.013	\\
Expected statistical uncertainty                           		&     0.124    	&         0.060    	\\
\hline
\end {tabular}
\end {center}
}
}
\vskip-0.2in
\caption {\label{tab:systematics}
Sources of systematic uncertainties, their one standard deviation variation level,  and their impact on fitting oscillation parameters.}
\end {table}

The effect of systematic uncertainties on the measured oscillation parameters was determined using Monte Carlo simulations in which modeling parameters were varied. 
Table~\ref{tab:systematics} shows the systematic effects, their 1\,$\sigma$ variation level,  and the impact on the values of mass splitting and mixing angle. 
Uncertainties in the physics simulations, including pion absorption cross-sections in the nucleus and associated modeling of energy deposition in the detector, result in the uncertainty in the visible hadronic energy (a), which is energy dependent and is about 7.0\% below \unit{3}\,GeV.  
The errors in the measurement of muon energy (b) from range (2\% error) or from curvature in the magnetic field (3\%) are included. 
The effects of relative reconstruction efficiencies between the two detectors and uncertainties in their fiducial masses and relative difference in detector structure result in the 1.6\% normalization error (c). 
These three uncertainties dominate the systematic error on the neutrino mass splitting. 
The largest uncertainty in the mixing angle is from the amount of neutral current background (d), the uncertainty on which, based on a data-driven method, is 20\%~\cite{ref:JasminePhD}. 
Other sources of uncertainty include: the 2.2\% relative energy calibration uncertainty between the two detectors (e);  uncertainties in the neutrino cross-sections $\sigma_\nu$ (f); the beam flux (g); and uncertainties due to misclassification of  neutrino and antineutrino interactions (h). 
Finally, incorporation of partially reconstructed events introduces a small uncertainty due to approximations made in modeling the rock composition and details of the Far Detector's edges (i). 
%

\begin{figure}[t]
\begin{center}
\includegraphics[viewport= 0 15 567 603, keepaspectratio,width= 0.49 \textwidth, clip=true]{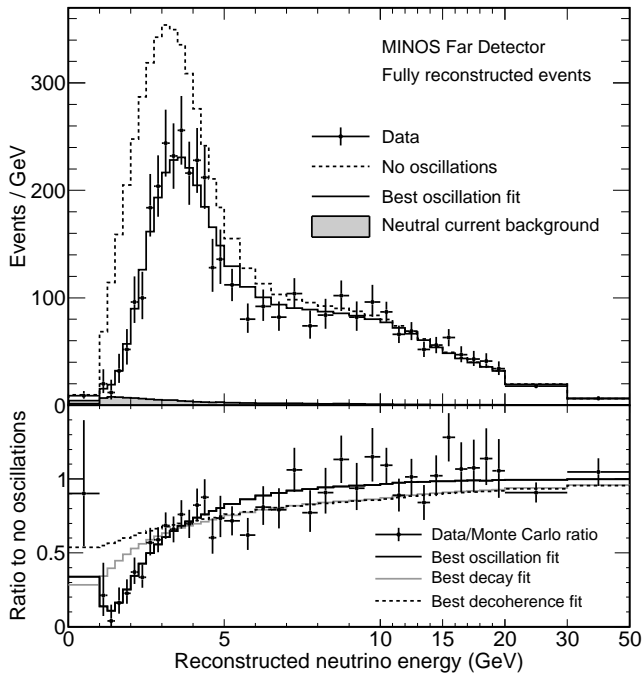}
\end{center}
\vskip-0.15in
\caption {
Top: The energy spectra of  fully reconstructed events in the Far Detector classified as charged current interactions.
The dashed histogram represents the spectrum predicted from measurements in  the Near Detector assuming no oscillations, while the solid histogram reflects the best fit of the oscillation hypothesis. The shaded area shows the predicted neutral current background.
Bottom: The points with error bars are the background-subtracted ratios of data to the no-oscillation hypothesis. 
Lines show the best fits for: oscillations,  decay~\cite{ref:neutrino-decay}, and decoherence~\cite{ref:decoherence}.
}
\label{fig:prediction-data}
\end{figure}


\begin{figure}[t]
\begin{center}
\includegraphics[viewport= 0 35 567 385,keepaspectratio,width= 0.49 \textwidth, clip=true]{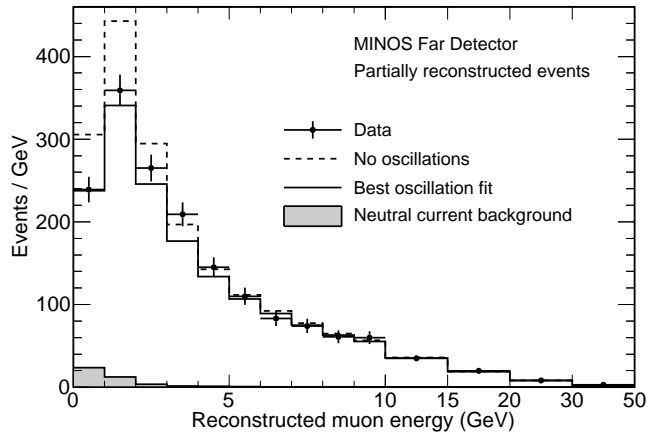}
\end{center}
\caption {The muon energy spectra of  partially reconstructed events in the Far Detector.
Conventions as in Fig.~\ref{fig:prediction-data}. }
\label{fig:prediction-RAFdata}
\end{figure}


\begin {table}[b]
{
{
\begin {center}
\begin {tabular} {c | c c  c | c | c c }
\hline
Run 			&   ~~POT~~ 		&
\multicolumn{2}{ | c } {Predicted}  &   \multicolumn{2}  {| c} {Observed} \\
Period		& $(10^{20})$	&
\multicolumn{2}{ | c } {(No oscillations)}  &   \multicolumn{2}  {| c} {(Far Detector)} \\
			    &
			& \multicolumn{1} {| c } {~~~Fully} & \multicolumn{1} { c } {~~~Partially} 
			& \multicolumn{1} {| c } {~~~Fully} & \multicolumn{1} { c } {~~~Partially} \\
			
\hline
I			& 1.269		
			& \multicolumn{1} {| r } {426 } & \multicolumn{1} { r } {375 } 
		     	& \multicolumn{1} {| r } {318 } & \multicolumn{1} { r } {357 } \\

II		     	& 1.943
			& \multicolumn{1} {| r } {639 } & \multicolumn{1} { r } {565 } 
		    	& \multicolumn{1} {| r } {511 } & \multicolumn{1} { r } {555 } \\

III		     	& 3.881
			& \multicolumn{1} {| r } {1,252 } & \multicolumn{1} { r } {1,130 } 
		     	& \multicolumn{1} {| r } {1,037 } & \multicolumn{1} { r } {977 } \\

HE			& 0.153
			& \multicolumn{1} {| r } {134 } & \multicolumn{1} { r } {136 } 
		     	& \multicolumn{1} {| r } {120 } & \multicolumn{1} { r } {128 } \\
\hline
Total			& 7.246
			& \multicolumn{1} {| r } {2,451 } & \multicolumn{1} { r } {2,206 } 
		     	& \multicolumn{1} {| r } {1,986 } & \multicolumn{1} { r } {2,017 } \\

\hline
\end {tabular}
\end {center}
}
}
\vskip-0.1in
\caption {\label{tab:events}
Numbers of events classified in the Far Detector as fully and
partially reconstructed charged current interactions shown for all running periods.
The predicted numbers are calculated under the assumption of no oscillations.
 }
\end {table}

All event selection criteria and analysis procedures were defined prior to examining the full data set in the Far Detector.
The energy spectra were compared with those used in the previous publication~\cite{ref:MINOS}.
These agree within the small differences expected due to changes in the reconstruction algorithm.  
The observed and predicted numbers of events, classified in the Far Detector as fully and partially reconstructed charged current interactions, for all running periods are shown in Table~\ref{tab:events}.
The energy spectrum of the fully reconstructed Far Detector data sample is shown in Fig.~\ref{fig:prediction-data}, along with the predicted spectra.
The corresponding spectra for the partially reconstructed events are shown in Fig.~\ref{fig:prediction-RAFdata}.

\begin{figure}[t]
\begin{center}
\includegraphics[viewport=10 70  547 507, keepaspectratio,width= 0.49 \textwidth, clip=true]{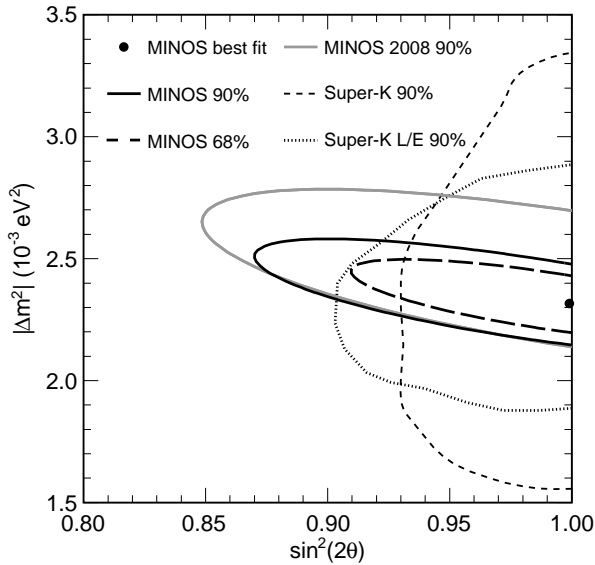}
\end{center}
\caption{Likelihood contours of 68\% and 90\%\,C.L. around the best fit values  for the mass splitting and mixing angle.
Also shown are contours from previous measurements~\cite{ref:SuperK,ref:MINOS}. 
}
\label{fig:contours}
\end{figure}

To test the neutrino oscillation model against the data, the two-parameter survival probability formula
$ P(\nu_\mu \rightarrow \nu_\mu) = 1 - \sin^2(2\theta) \sin^2 ({\Delta m^2 L / 4 E})$
was used.
The best values of $|\Delta m^2|$ and $\rm sin^2\!(2\theta)$ were found by maximizing a likelihood, which includes the four dominant systematic uncertainties a--d in Table~\ref{tab:systematics} as nuisance fit parameters~\cite{Rolke:2004mj,ref:JessMitchell}.
The likelihood value is computed at each point in the $|\Delta m^2| - \rm sin^2\!(2\theta)$ plane by summing the contributions from the seven event categories.   
Within each category the likelihood value is calculated by comparing the observed energy spectrum with that predicted for the oscillation parameters of that point.
The best fit value and one-dimensional uncertainties for the mass splitting are $|\Delta m^2| = (2.32^{+0.12}_{-0.08})\times10^{-3}$\,eV$^2$.
For the mixing angle, if $\rm sin^2\!(2\theta)$ is constrained  to be  $ \le 1$, the best fit is $\rm sin^2\!(2\theta)=1.00$ or  $\rm sin^2\!(2\theta)  > 0.94 ~(0.90)$  at 68~(90)\% confidence level (C.L.).
The best fit values with the resulting 68\% and 90\%\,C.L. contours are shown in Fig.~\ref{fig:contours}.
Imposition of the physical boundary on $\rm sin^2\!(2\theta)$ results in the total uncertainty being smaller than that expected from the simulated sensitivity.
Without this constraint the best fit value for mass splitting changes by  $\delta(\Delta m^2) = -0.01\times10^{-3}$\,eV$^2$  and for the mixing angle changes by $\delta(\rm sin^2\!(2\theta))=+0.001$. 
The fits do not significantly pull away from their nominal values any of the four nuisance parameters. 
Predicted energy spectra for the best fit are shown in Fig.~\ref{fig:prediction-data}.
If the fit is restricted to use only fully reconstructed  events with the negative track charge,  the best fit value for mass splitting changes by  $\delta(\Delta m^2) = +0.03\times10^{-3}$\,eV$^2$ and the mixing angle is unchanged.
Two other hypotheses for neutrino disappearance, pure neutrino decay~\cite{ref:neutrino-decay} and pure quantum decoherence~\cite{ref:decoherence}, are excluded at $7$ and $9$ standard deviations, respectively, as shown in Fig.~\ref{fig:prediction-data}.

In summary,  MINOS data from a beam exposure of $7.25 \times 10^{20}$\,POT, more than double the data set used in the previous MINOS publication~\cite{ref:MINOS}, and improved analysis methodology have resulted in the measurement of the value of  the atmospheric mass splitting to be $|\Delta m^2| =  $\unit{$(2.32^{+0.12}_{-0.08})\times10^{-3}$}\,eV$^2$ and the mixing angle to be $\rm sin^2\!(2\theta)  > 0.90$ (90\%\,C.L.).
This is the most precise  measurement of this mass splitting to date.
Neither the pure quantum decoherence nor neutrino decay hypotheses fit the observed spectra.


This work was supported by the US DOE; the UK STFC; the US NSF; the State and University of Minnesota; the University of Athens, Greece; and Brazil's FAPESP, CNPq, and CAPES.  
We are grateful to the Minnesota DNR, the crew of the Soudan Underground Laboratory, and the personnel of Fermilab for their contributions to this effort.



\end{document}